\documentclass[seceq]{ptptex}
\usepackage{wrapft}

\newcommand{\id}{{1\!\!1}} 
\newcommand {\be}{\begin{equation}}
\newcommand {\ee}{\end{equation}}
\newcommand {\bea}{\begin{eqnarray}}
\newcommand {\eea}{\end{eqnarray}}
\newcommand {\nn}{\nonumber}
\newcommand {\tr}{{\rm tr\,}}
\newcommand {\Tr}{\mbox{Tr\,}}
\newcommand {\ket}{\rangle}
\newcommand {\bra}{\langle}

\newcommand {\bA}{{\tt A}}
\newcommand {\bB}{{\tt B}}
\newcommand {\bC}{{\tt C}}
\newcommand {\cA}{{\cal A}}
\newcommand {\cN}{{\cal N}}
\newcommand {\cF}{{\cal F}}
\newcommand {\cO}{{\cal O}}
\newcommand {\bq}{{\bf q}}

\preprintnumber[4cm]{
WIS/05/10-APR-DPPA \\
OIQP-10-03}

\title{
Two-dimensional lattice for four-dimensional 
${\cal N}=4$ supersymmetric Yang-Mills
}

\author{
Masanori \textsc{Hanada}$^{1,2,}$\footnote{E-mail: mhanada@u.washington.edu}, 
So \textsc{Matsuura}$^{3,}$\footnote{E-mail: s.matsu@phys-h.keio.ac.jp} 
and Fumihiko \textsc{Sugino}$^{4,}$\footnote{E-mail: fumihiko\_sugino@pref.okayama.lg.jp}%
}

\inst{
$^1$Department of Particle Physics and Astrophysics, Weizmann Institute of Science,
Rehovot 76100, Israel\\
$^2$Department of Physics, University of Washington, Seattle, WA 98195-1560, USA\\
$^3$Department of Physics, 
and Research and Education Center for Natural Science, 
Keio University, 4-1-1 Hiyoshi, Yokohama, 223-8521, Japan\\
$^4$
Okayama Institute for Quantum Physics, Kyoyama 1-9-1, Kita-ku, Okayama 700-0015, Japan
}

\abst{
We construct a lattice formulation of a mass-deformed 
two-dimensional $\cN=(8,8)$ super Yang-Mills theory  
with preserving two supercharges exactly. 
Gauge fields are represented by compact unitary link variables, 
and the exact supercharges on the lattice are nilpotent up to gauge transformations and 
$SU(2)_R$ rotations.  
Due to the mass deformation, the lattice model is free from the vacuum degeneracy problem, 
which was encountered in earlier approaches, 
and flat directions of scalar fields are stabilized 
giving discrete minima representing fuzzy $S^2$. 
Around the trivial minimum, quantum continuum theory is obtained with no tuning, 
which serves a nonperturbative construction of the IIA matrix string theory. 
Moreover, around the minimum of $k$-coincident fuzzy spheres, four-dimensional $\cN=4$ $U(k)$ super Yang-Mills 
theory with two commutative and two noncommutative directions emerges. 
In this theory, sixteen supersymmetries are broken by the mass deformation to two. 
Assuming the breaking is soft, we give a scenario leading to undeformed $\cN=4$ super Yang-Mills on 
${\bf R}^4$ without any fine tuning. 
As an evidence for the validity of the assumption, some computation of 1-loop radiative 
corrections is presented.  
}

\begin{document}

\maketitle

\section{Introduction} 
Nonperturbative aspects of 
supersymmetric Yang-Mills (SYM) theories play prominent roles in 
physics beyond the standard model~\cite{Witten:1981nf} as well as 
in superstring/M theory~\cite{BFSS,IKKT,MatrixString,Maldacena:1997re}. 
However, to construct their nonperturbative formulations such as lattice is not 
a straightforward task because of the notorious 
difficulties of supersymmetry (SUSY) on lattice. So far, lattice formulations for SYM 
are constructed 
for one- 
and two-dimensional 
cases and ${\cal N}=1$ pure SYM in three and four dimensions~\cite{Giedt:2009yd}, 
where no requirement of fine tunings due to the ultra-violet (UV) effects can be shown at least 
in perturbative arguments. 
For one-dimensional theory (matrix quantum mechanics) 
more powerful ``non-lattice'' 
technique~\cite{Hanada-Nishimura-Takeuchi} 
is applicable. 
(For corresponding lattice study, see Ref.~\citen{Catterall-Wiseman}.) 
For two-dimensional $\cN=(2,2)$ SYM, nonperturbative evidences for 
the lattice model presented in Ref.~\citen{Sugino:2004qd} to require no fine tuning 
have been given by numerical simulation for the gauge group $G=SU(2)$ in Ref.~\citen{Kanamori:2008bk} 
and for $G=SU(N)$ with $N=2,3,4,5$ in Ref.~\citen{Hanada:2009hq}~\footnote{Recently, 
Ref.~\citen{Hanada:2010qg} has shown that the model constructed in Ref.~\citen{kaplan_etal} is 
free from the sign problem and 
gives the same physics as that in Ref.~\citen{Sugino:2004qd} after an appropriate treatment of the overall 
$U(1)$ modes.}.   
Combining such techniques   
with the plane wave deformation \cite{BMN} and the Myers effect \cite{Myers:1999ps}, 
three-dimensional theory can be obtained 
as a theory on fuzzy sphere~\cite{Maldacena:2002rb}. 
Also, in the planar limit, four-dimensional theory 
can be obtained using a novel large-$N$ reduction technique~\cite{Ishii:2008ib} 
inspired by the Eguchi-Kawai equivalence~\cite{Eguchi:1982nm}. 
However, four-dimensional theories of extended SUSY at a finite rank of a gauge group 
are still out of reach.

We consider 
$\cN=(8,8)$ SYM with a mass deformation 
reminiscent of the ``plane wave matrix string''~\cite{Das:2003yq,Bonelli:2002mb} in the next section,  
and construct a lattice formulation of the theory preserving two supercharges~\footnote{
For simplicity, 
we focus on the gauge group $G = U(N)$, although the similar argument is valid also for $G=SU(N)$. } 
in section~\ref{sec:lattice}. 
It is a modification of a lattice formulation by one of the authors 
(F. S.)~\cite{Sugino:2003yb,Sugino:2004qd,Sugino:2004uv}. 
(For related constructions, 
see Refs.~\citen{Cohen:2003xe,Catterall:2004np,D'Adda:2005zk}.)  
Thanks to the deformation, it is free from the vacuum degeneracy problem encountered in 
Ref.~\citen{Sugino:2004uv} 
as well as from the problem of flat directions. 
In a perturbative argument, 
it is shown that the continuum theory is obtained  
without any fine tuning.  
If we turn off the mass parameter in the continuum theory, 
undeformed ${\cal N}=(8,8)$ SYM in two dimensions, i.e. 
the IIA matrix string theory~\cite{MatrixString} is obtained. 
In section~\ref{sec:4dN=4}, 
we furthermore present an intriguing scenario leading to four-dimensional $\cN=4$ SYM 
with a finite-rank 
gauge group $G=U(k)$, starting with the lattice formulation of the two-dimensional theory. 
In order to realize the four-dimensional theory, we consider a $k$-coincident 
fuzzy sphere solution in the mass-deformed two-dimensional theory 
and take a large-$N$ limit with scaling the deformation 
parameter appropriately. 
The crucial point is that one takes the continuum limit 
as two-dimensional theory first~\footnote{
Similar asymmetric continuum limit is discussed on four-dimensional lattice~\cite{Kaplan:2002wv} 
in order to reduce the number of fine tunings. 
} 
and then lifts the two-dimensional continuum theory to four dimensions using 
a matrix model formulation of noncommutative space. 
Section~\ref{sec:conclusion} is devoted to discuss possible future directions. 

\setcounter{equation}{0}
\section{2d continuum theory}
\label{sec:continuum}
We start with continuum $\cN=(8,8)$ SYM on ${\bf R}^2$: 
\begin{eqnarray}
S_{0}
&=&
\frac{1}{g_{2d}^2}\int d^2x\ {\rm Tr}
\biggl\{
F_{12}^2
+
\left(D_\mu X^I\right)^2
-
\frac{1}{2}\left[X^I,X^J\right]^2
\nonumber\\
& &
\hspace{24mm}
+
\Psi^T\left(D_1+\gamma_2D_2\right)\Psi
+
i\Psi^T\gamma_I\left[X^I,\Psi\right]
\biggl\},
\label{S0}
\end{eqnarray}
where 
$\mu=1,2$, $I=3,\cdots,10$, and $D_\mu = \partial_\mu +i[A_\mu, \cdot]$.    
$16\times 16$ gamma matrices $\gamma_i$ ($i=2,\cdots,10$)  
satisfy
$\{\gamma_i, \gamma_j\}=-2\delta_{ij}$.

We rewrite this action by using Hermitian scalars 
$X_i\ (i=3,4), B_{\bA}\ (\bA=1,2,3)$ and $C$, complex scalars 
$\phi_\pm$, 
bosonic auxiliary fields $H_{\bA}$, 
$\tilde{H}_\mu$, $\tilde{h}_i$,   
and fermionic variables $\psi_{\pm\mu}, \rho_{\pm i}, \chi_{\pm \bA}$ and $\eta_{\pm}$. 
As presented in appendix~\ref{app:notation}, 
scalars ($B_{\bA}, C, \phi_\pm$) and the fermionic variables are related to $X_I$ ($I=5,\cdots,10$) 
and spinor components of $\Psi$ by a simple rearrangement, respectively.  
There are appropriate supercharges $Q^{(0)}_{\pm}$ by which
$S_0$ can be written in exact form~\footnote{
This is obtained from BTFT formulation of four-dimensional ${\cal N}=4$ SYM in Ref.~\citen{Sugino:2003yb} by 
dimensional reduction. Here, we redefine $H_{\bA}+\frac12\epsilon_{\bA\bB\bC}[B_{\bB}, B_{\bC}]$, 
$\phi$, $\bar{\phi}$  
in (4.13) in Ref.~\citen{Sugino:2003yb} as $H_{\bA}$, $\phi_+$, $\phi_-$, respectively.   
} as   
\begin{eqnarray} 
S_0
=
Q^{(0)}_+Q^{(0)}_-
{\cal F}^{(0)}, 
\label{Q^{(0)}-closed form}
\end{eqnarray}
where 
\begin{eqnarray}
{\cal F}^{(0)}
&=&
\frac{1}{g^2_{2d}}\int d^2x\ {\rm Tr}\Bigl\{
-iB_{\bA}\Phi_{\bA}
-
\frac{1}{3}\epsilon_{\bA\bB\bC}B_{\bA}[B_{\bB},B_{\bC}]
\nonumber\\
& &
\hspace{24mm}
-
\psi_{+\mu}\psi_{-\mu}
-
\rho_{+i}\rho_{-i}
-
\chi_{+\bA}\chi_{-\bA}
-
\frac{1}{4}\eta_+\eta_-
\Bigl\},  
\label{F}
\label{4d N=4 continuum action}
\end{eqnarray}
and $\Phi_1=2(-D_1X_3-D_2X_4)$, 
$\Phi_2=2(-D_1X_4+D_2X_3)$,  
$\Phi_3=2(-F_{12}+i[X_3,X_4])$. 
Supercharges $Q_{\pm}^{(0)}$ transform fields as  
\begin{eqnarray}
& &
Q^{(0)}_{\pm}A_\mu=\psi_{\pm\mu},
\quad
Q_\pm\psi_{\pm\mu}
=
\pm iD_\mu\phi_{\pm}, 
\quad Q^{(0)}_\mp\psi_{\pm\mu}
=
\frac{i}{2}D_\mu C\mp\tilde{H}_\mu, 
\nonumber\\
& &
Q^{(0)}_{\pm}\tilde{H}_\mu
=
\left[\phi_{\pm},\psi_{\mp\mu}\right]
\mp\frac{1}{2}\left[C,\psi_{\pm\mu}\right]
\mp\frac{i}{2}D_\mu\eta_{\pm},
\nonumber\\ 
& &
Q^{(0)}_{\pm}X_i=\rho_{\pm i},
\quad
Q^{(0)}_\pm\rho_{\pm i}
=
\mp \left[X_i,\phi_{\pm}\right], 
\quad 
Q^{(0)}_\mp\rho_{\pm i}
=
-\frac{1}{2}[X_i, C]\mp\tilde{h}_i, 
\nonumber\\
& &
Q^{(0)}_{\pm}\tilde{h}_i
=
\left[\phi_{\pm},\rho_{\mp i}\right]
\mp\frac{1}{2}\left[C,\rho_{\pm i}\right]
\pm\frac{1}{2}\left[X_i,\eta_{\pm}\right], 
\nonumber\\
& &
Q^{(0)}_{\pm}B_{\bA}
=
\chi_{\pm \bA}, 
\quad
Q^{(0)}_{\pm}\chi_{\pm \bA}
=
\pm[\phi_\pm,B_{\bA}], 
\quad 
Q^{(0)}_\mp\chi_{\pm A}
=
-\frac{1}{2}[B_{\bA},C]
\mp H_{\bA}, 
\nonumber\\
& &
Q^{(0)}_\pm H_{\bA}
=
[\phi_\pm,\chi_{\mp \bA}]
\pm\frac{1}{2}\left[B_{\bA},\eta_\pm\right]
\mp\frac{1}{2}\left[C,\chi_{\pm \bA}\right], 
\nonumber\\ 
& &
Q^{(0)}_\pm C
=
\eta_\pm, 
\quad
Q^{(0)}_\pm\eta_\pm
=
\pm\left[\phi_\pm,C\right], 
\quad
Q^{(0)}_\mp\eta_\pm
=
\mp\left[\phi_+,\phi_-\right], 
\nonumber\\
& &
Q^{(0)}_\pm\phi_\pm=0, 
\quad
Q^{(0)}_{\mp}\phi_\pm=\mp\eta_\pm.  
\label{SUSY algebra_4dN=4_continuum}
\end{eqnarray}
One can see the nilpotency 
$\left(Q_+^{(0)}\right)^2 = \left(Q_-^{(0)}\right)^2 
= \{Q_+^{(0)},Q_-^{(0)}\}=0$  
up to gauge transformations. 

\subsection{Mass deformation} 
Next, we introduce a mass $M$ to deform these charges as 
\begin{eqnarray}
Q_{\pm}=Q_\pm^{(0)}+\Delta Q_{\pm}, 
\end{eqnarray}
where non-vanishing $\Delta Q_\pm$ transformations are  
\begin{eqnarray}
& & 
\Delta Q_{\pm}\tilde{H}_\mu
= 
\frac{M}{3}\psi_{\pm\mu},
\qquad
\Delta Q_{\pm}\tilde{h}_i
= 
\frac{M}{3}\rho_{\pm i}, 
\qquad 
\Delta Q_\pm H_{\bA}
= 
\frac{M}{3}\chi_{\pm \bA}, 
\nonumber \\
& & \Delta Q_\pm\eta_\pm
= 
\frac{2M}{3}\phi_\pm, 
\qquad 
\Delta Q_\mp\eta_\pm
= 
\pm\frac{M}{3}C. 
\label{SUSY algebra_deformation}
\end{eqnarray}
Then $Q_\pm$ satisfy the anti-commutation relations,  
\begin{equation}
Q_+^2
=
\frac{M}{3}J_{++}, 
\quad
Q_-^2
=
-\frac{M}{3}J_{--}, 
\quad 
\{Q_+,Q_-\}
=
-\frac{M}{3}J_0,  
\label{Q-anticommutator}
\end{equation}
up to gauge transformations, where $J_0$, $J_{++}$ and $J_{--}$ are 
generators of $SU(2)_R$ symmetry acting to fields as 
\bea
J_{++} & = & \int d^2x \left[\psi_{+\mu}^{\alpha}(x)\frac{\delta}{\delta\psi_{-\mu}^{\alpha}(x)} 
+\chi_{+\bA}^{\alpha}(x)\frac{\delta}{\delta\chi_{-\bA}^{\alpha}(x)} 
-\eta_+^{\alpha}(x)\frac{\delta}{\delta\eta_-^{\alpha}(x)} \right. \nn \\
 & & \hspace{14mm}
\left. +2\phi_+^{\alpha}(x) \frac{\delta}{\delta C^{\alpha}(x)} 
-C^{\alpha}(x) \frac{\delta}{\delta\phi_-^{\alpha}(x)}\right], \nn \\
J_{--} & = & \int d^2x \left[\psi_{-\mu}^{\alpha}(x)\frac{\delta}{\delta\psi_{+\mu}^{\alpha}(x)} 
+\chi_{-\bA}^{\alpha}(x)\frac{\delta}{\delta\chi_{+\bA}^{\alpha}(x)} 
-\eta_-^{\alpha}(x)\frac{\delta}{\delta\eta_+^{\alpha}(x)} \right. \nn \\
 & & \hspace{14mm}
\left. -2\phi_-^{\alpha}(x) \frac{\delta}{\delta C^{\alpha}(x)} 
+C^{\alpha}(x) \frac{\delta}{\delta\phi_+^{\alpha}(x)}\right], \nn \\
J_0 & = & \int d^2x \left[ \psi_{+\mu}^{\alpha}(x)\frac{\delta}{\delta\psi_{+\mu}^{\alpha}(x)} 
-\psi_{-\mu}^{\alpha}(x)\frac{\delta}{\delta\psi_{-\mu}^{\alpha}(x)} \right. \nn \\
 & & \hspace{14mm}
+\chi_{+\bA}^{\alpha}(x)\frac{\delta}{\delta\chi_{+\bA}^{\alpha}(x)} 
-\chi_{-\bA}^{\alpha}(x)\frac{\delta}{\delta\chi_{-\bA}^{\alpha}(x)}  
+\eta_+^{\alpha}(x)\frac{\delta}{\delta\eta_+^{\alpha}(x)} 
-\eta_-^{\alpha}(x)\frac{\delta}{\delta\eta_-^{\alpha}(x)}   \nn \\
 & & \hspace{14mm}
\left. 
 +2\phi_+^{\alpha}(x) \frac{\delta}{\delta \phi_+^{\alpha}(x)} 
-2\phi_-^{\alpha}(x) \frac{\delta}{\delta\phi_-^{\alpha}(x)}\right].  
\eea  
($\alpha$ is an index for the gauge group generators.)   
The eigenvalues of $J_0$ are $\pm 1$ for the fermions with index $\pm$, 
$\pm 2$ for $\phi_\pm$, and zero for the other bosonic fields. 
Note that $\phi_\pm$ and $C$ form an $SU(2)_R$ triplet and each pair of 
$(\psi_{+\mu},\psi_{-\mu})$, 
$(\chi_{+\bA},\chi_{-\bA})$, 
$(\eta_+,-\eta_-)$ and $(Q_+,Q_-)$ forms a doublet.   
In particular, 
\begin{equation}
[J_{\pm\pm}, Q_\pm]=0, \qquad [J_{\pm\pm}, Q_\mp]=Q_\pm, \qquad [J_0,Q_\pm] =\pm Q_\pm . 
\label{Q_doublet}
\end{equation}
Using the modified supercharges, we can define $Q_\pm$-closed action as~\footnote{
This kind of deformation is extended to various SYM theories in Ref.~\citen{Kato:2011yh}.}  
\begin{eqnarray}
S=
\left(
Q_+Q_- - \frac{M}{3}
\right)
{\cal F}, 
\label{Q-closed form}
\end{eqnarray}
where 
\be
{\cal F}={\cal F}^{(0)}+\Delta {\cal F}, 
\qquad 
\Delta {\cal F}=\frac{1}{g_{2d}^2}\int d^2x\ 
{\rm Tr}\left(
\sum_{\bA=1}^3 \frac{a_{\bA}}{2}B_{\bA}^2
+
\sum_{i=3}^4
\frac{c_i}{2}X_i^2
\right). 
\label{cF_cont}
\ee
That the action \eqref{Q-closed form} is $Q_\pm$-closed can easily be seen 
by using \eqref{Q-anticommutator}, (\ref{Q_doublet}) 
and $SU(2)_R$ invariance of ${\cal F}$. 
After integrating out the auxiliary fields, 
$B_{\bA}$ and $X_i$ have positive mass terms 
as long as the parameters $a_{\bA}$ and $c_i$ all lie  
in the open interval $(-2M/3,0)$. 
Here we take 
$a_1=a_2=a_3=-\frac{2M}{9}$ and 
$c_3=c_4=-\frac{4M}{9}$ 
for convenience. 
Then the action reads 
\begin{eqnarray}
S=S_0+\Delta S, 
\end{eqnarray}
where
\begin{eqnarray}
& &
\Delta S
=
\frac{1}{g_{2d}^2}\int d^2x\ {\rm Tr}\Bigl\{
\frac{2M^2}{81}\left(
B_{\bA}^2 + X_i^2
\right)
+
\frac{M^2}{9}
\left(
\frac{C^2}{4}+\phi_+\phi_-
\right)
-\frac{M}{2}C[\phi_+,\phi_-]
\nonumber\\ 
& &
\hspace{34mm}
+
\frac{2M}{3}\psi_{+\mu}\psi_{-\mu}
+
\frac{2M}{9}\rho_{+i}\rho_{-i}
+
\frac{4M}{9}\chi_{+A}\chi_{-A}
-
\frac{M}{6}\eta_+\eta_-
\nonumber\\
& & 
\hspace{34mm}
-\frac{4iM}{9}B_3\left(
F_{12}+i[X_3,X_4]
\right)
\Bigl\}. 
\label{DeltaS}
\end{eqnarray}
{}From this expression one can see some similarity 
to the plane wave matrix model~\cite{BMN} and to PP wave matrix strings~\cite{Das:2003yq}. 
The first two terms in the first line and the terms in the second line in (\ref{DeltaS}) 
give mass terms to scalars and to fermions, respectively. 
The third term represents the so called Myers term \cite{Myers:1999ps}. 
Thanks to these terms, 
fuzzy $S^2$ configurations satisfying 
\be 
[\phi_+,\phi_-]=\frac{M}{3}C, \qquad [C, \phi_\pm] = \pm \frac{2M}{3} \phi_\pm , 
\qquad B_{\bA} = X_i =0
\label{fuzzy_S2_eq}
\ee
give the minima of the action ($S=0$) preserving $Q_\pm$ SUSYs. 
Note that the last term in (\ref{DeltaS}) is purely imaginary. 
Also, we should recognize that the mass-deformed action preserves only two supercharges ($Q_\pm$) but 
the other 14 charges are softly broken by the deformation. 


\setcounter{equation}{0}
\section{Lattice formulation}
\label{sec:lattice}
In this section we put the deformed theory on a two-dimensional square lattice. 
We use link variables $U_\mu=e^{iaA_\mu}$ belonging to the gauge group $U(N)$, 
where $a$ is the lattice spacing. 
Other lattice fields, defined on sites, are made dimensionless by multiplying 
suitable powers of $a$ to the continuum counterparts: 
\bea
& & \mbox{(scalars)}^{\rm lat} = a \,\mbox{(scalars)}^{\rm cont}, \qquad 
\mbox{(fermions)}^{\rm lat} =a^{3/2}\mbox{(fermions)}^{\rm cont}, \nn \\
& & Q_\pm^{\rm lat} = a^{1/2}Q_{\pm}^{\rm cont}.  
\eea
Also, dimensionless coupling constants on the lattice are  
\be
g_0=a g_{2d}, \qquad M_0=a M.
\ee 
The supersymmetry transformations are realized as 
\begin{eqnarray}
& & \hspace{-4mm}
Q_\pm U_\mu(x)=i\psi_{\pm\mu}(x)U_\mu(x), 
\nonumber\\
& & \hspace{-4mm}
Q_\pm\psi_{\pm\mu}(x)
=
i\psi_{\pm\mu}(x)\psi_{\pm\mu}(x)
\pm i D_\mu \phi_\pm(x), 
\nonumber\\
& & \hspace{-4mm}
Q_\mp\psi_{\pm\mu}(x)
=
\frac{i}{2}\left\{\psi_{+\mu}(x),\psi_{-\mu}(x)\right\}
+\frac{i}{2} D_\mu C(x)
\mp\tilde{H}_\mu(x), 
\nonumber\\
& & \hspace{-4mm}
Q_\pm\tilde{H}_\mu(x)
=
-\frac{1}{2}\left[
\psi_{\mp\mu}(x),\phi_{\pm}(x)+U_\mu(x)\phi_\pm(x+\hat{\mu})U_\mu(x)^\dagger
\right]
\nonumber\\
& &
\hspace{1.7cm}
\pm\frac{1}{4}\left[
\psi_{\pm\mu}(x),C(x)+U_\mu(x)C(x+\hat{\mu})U_\mu(x)^\dagger
\right]
\nonumber\\ 
& &
\hspace{1.7cm}
\mp\frac{i}{2} D_\mu \eta_\pm(x)
\pm\frac{1}{4}\left[
\psi_{\pm\mu}(x)\psi_{\pm\mu}(x),\psi_{\mp\mu}(x)
\right]
\nonumber\\
& &
\hspace{1.7cm}
+
\frac{i}{2}\left[
\psi_{\pm\mu}(x),\tilde{H}_\mu(x)
\right]
+
\frac{M_0}{3}\psi_{\pm\mu}, 
\label{SUSY on lattice}
\end{eqnarray}
for the lattice fields
$U_\mu, \psi_{\pm\mu}$ and $\tilde{H}_\mu$. 
$D_\mu$ is a covariant forward difference operator defined by 
\begin{equation}
D_\mu A(x) \equiv U_\mu(x) A(x+\hat{\mu}) U_\mu(x)^\dagger -A(x), 
\end{equation}
for any adjoint field $A(x)$. 
Transformation of the other fields is  
the same as the one in continuum with the obvious replacement $M \to M_0$. 
Then the anti-commutation relation \eqref{Q-anticommutator} holds on the lattice with $M \to M_0$. 
In order to construct a corresponding lattice action, we take lattice counterparts of $\Phi_{\bA}$ as  
\begin{eqnarray}
& & \hspace{-1cm}\Phi_1(x)
=
2\left(-D_1X_3(x) -D_2X_4(x)\right), 
\nonumber\\
& & \hspace{-1cm}\Phi_2(x)
=
2\left(-D_1^*X_4(x) +D_2^*X_3(x)\right), 
\nonumber\\
& & \hspace{-1cm}\Phi_3(x)
=
\frac{i(U_{12}(x)-U_{21}(x))}{1-\epsilon^{-2}||1-U_{12}(x)||^2}
+
2i[X_3(x),X_4(x)], 
\label{Phi3_lat}
\end{eqnarray}
where $D_\mu^*$ is a covariant backward difference operator,   
\begin{equation}
D_\mu^* A(x) \equiv A(x) -U_\mu(x-\hat{\mu})^\dagger A(x-\hat{\mu}) U_\mu(x-\hat{\mu}), 
\end{equation}
$U_{\mu\nu}(x)=U_\mu(x)U_\nu(x+\hat{\mu})U_\mu(x+\hat{\nu})^\dagger U_\nu(x)^\dagger$ 
is a plaquette variable, 
$\epsilon$ is a positive constant satisfying $0<\epsilon<2$,   
and the norm of a matrix is defined by $||A||=\sqrt{{\rm Tr}(AA^\dagger)}$. 
The first term of the r.h.s. of $\Phi_3(x)$ is a lattice counterpart of the field strength $F_{12}$. 
It is the same as 
the situation in the lattice formulation for two-dimensional $\cN=(2,2)$ $U(N)$ SYM in 
Ref.~\citen{Sugino:2004qd}. 
$Q_\pm$-invariant lattice action is given as 
\be
S_{\rm lat} = \left(Q_+Q_--\frac{M_0}{3}\right)\cF_{\rm lat}
\ee
with $\cF_{\rm lat}$ being the same form as $\cF$ in (\ref{cF_cont}) under the trivial replacement 
$\frac{1}{g_{2d}^2}\int d^2x \to \frac{1}{g_0^2}\sum_x$, $M \to M_0$, 
when the admissibility condition $||1-U_{12}(x)||<\epsilon$ is satisfied for $\forall x$. 
Otherwise, $S_{\rm lat}=+\infty$. 

Note that in Eqs.~(\ref{Phi3_lat}) the covariant forward difference is used for $\Phi_1$, while 
the covariant backward difference is used for $\Phi_2$. 
With this choice, no species doubler appears in both of bosonic and fermionic kinetic terms.  
Note also that the fuzzy sphere solution of the lattice version of the equations (\ref{fuzzy_S2_eq}) 
\be
C=\frac{2M_0}{3}L_3, 
\qquad
\phi_\pm=\frac{M_0}{3}(L_1 \pm iL_2),  \qquad B_{\bA} = X_i =0
\label{fuzzy sphere solution}
\ee
preserves 
$Q_\pm$ SUSYs at regularized level, where 
$L_a$ ($a=1,2,3$) belong to an $N$-dimensional representation of $SU(2)$ generators satisfying  
$[L_a,L_b]=i\epsilon_{abc}L_c$.

\subsection{No unphysical degenerate minima} 
Here, we will check that the lattice action has the minimum only at the pure gauge configuration 
$U_{12}(x)=\id_N$, which guarantees that the weak field expansion 
$
U_{\mu}(x) = 1+ iaA_\mu(x) + \frac{(ia)^2}{2!}A_\mu(x)^2 + \cdots
$ 
is allowed in the continuum limit so that the lattice theory converges to the desired continuum theory 
at the classical level. 

After integrating out the auxiliary fields, bosonic part of the action $S_{\rm lat}$ takes the form  
\be
S_{\rm lat}^{(B)} = \frac{1}{g_0^2}\sum_x \tr \left[
\frac{2M_0^2}{81}\left(X_i(x)^2 + B_{\bA}(x)^2\right)-i\frac{2M_0}{9}B_3(x)\Phi_3^{(-)}(x)\right] +S_{\rm PDT} 
\label{S_lat_B}
\ee
where $\Phi_3^{(-)}(x)$ is $\Phi_3(x)$ in (\ref{Phi3_lat}) with the sign of the first term flipped, and 
$S_{\rm PDT}$ denotes  positive (semi-)definite terms. 
We will treat the second term, which is purely imaginary, as an operator in the reweighting method, and 
consider the minimum of the remaining part of $S_{\rm lat}^{(B)}$. 
The mass terms in (\ref{S_lat_B}) fix the minimum at 
\be
X_i(x)=B_{\bA}(x)=0,
\label{min_XB}
\ee 
which is independent of $S_{\rm PDT}$. At the minimum (\ref{min_XB}), $S_{\rm PDT}$ becomes 
\bea
S_{\rm PDT}& = & \frac{1}{g_0^2}\sum_x \tr \left[
\sum_{\mu} \left(D_\mu X_p(x)\right)^2 
+ \left(i[X_p(x), X_q (x)]+\frac{M_0}{3}\epsilon_{pqr} X_r(x) \right)^2\right] \nn \\
& & +\frac{1}{4g_0^2} \sum_x \frac{{\rm tr} \left[-(U_{12}(x)-U_{21}(x))^2\right]}{\left(1-\frac{1}{\epsilon^{2}}||1-U_{12}(x)||^2\right)^2}  
\eea 
with 
$C=2X_8$, $\phi_\pm = X_9 \pm iX_{10}$ and $p, q, r=8,9,10$. 
Since the last term representing the gauge kinetic term is the same  
as in the case of two-dimensional $\cN=(2,2)$ SYM discussed in Ref.~\citen{Sugino:2004qd}, 
the admissibility condition with $0<\epsilon<2$ for the gauge group $U(N)$ singles out 
the trivial minimum $U_{12}(x)=\id_N$. 
It shows that the lattice action has a stable physical vacuum and unphysical degeneracies of vacua seen 
in the former formulation \cite{Sugino:2004uv} do not appear.   
The mass deformation preserving $Q_\pm$ SUSYs is crucial 
to stabilize flat directions of scalars as well as to remove the unphysical minima for gauge fields.  

\subsection{No need of fine tuning} 
Next, we discuss in the perturbation theory that 
the desired quantum continuum theory is obtained  
without any fine tuning. 

In the theory near the continuum limit with the auxiliary fields integrated out, let us consider 
local operators of the type: 
\be
\cO_p(x) = M^m \varphi(x)^\alpha\partial^\beta \psi(x)^{2\gamma}, \qquad p\equiv m+\alpha+\beta+3\gamma
\ee
where $\varphi$ denotes scalar fields as well as gauge fields, $\psi$ fermionic fields, 
and $\partial$ derivatives. The mass dimension of ${\cal O}_p$ is $p$, 
and $m, \alpha, \beta, \gamma = 0,1,2, \cdots$.  
{}From dimensional analysis, it can be seen that  
radiative corrections from UV region of loop momenta to ${\cal O}_p$
have the form 
\be
\left(\frac{1}{g_{2d}^2}c_0 a^{p-4} + c_1 a^{p-2} +
g_{2d}^2 c_2 a^p +\cdots\right) \int d^2x \,\cO_p(x), 
\label{renormalization}
\ee
up to possible powers of $\ln(aM)$. $c_0, c_1, c_2$ are dimensionless numerical constants. 
The first, second and third terms in the parenthesis are contributions from tree, 1-loop and 2-loop effects, 
respectively. The ``$\cdots$'' is effects from higher loops, which are irrelevant for the analysis. 

Since relevant or marginal operators generated by loop effects possibly appear from nonpositive powers of $a$ 
in the second and third terms in (\ref{renormalization}), we should see operators with $p=0,1,2$. 
They are $\varphi$, $M\varphi$ and $\varphi^2$. (Note that $\id$, $M$, $M^2$ and $\partial\varphi$ are not 
dynamical.)
Candidates for $\varphi$ are linear combinations of $\tr X_i$ and $\tr B_{\bA}$ from gauge and $SU(2)_R$ 
symmetries. But, all of them are not invariant under $Q_\pm$ SUSYs, and thus are forbidden to 
appear. Similarly, $M\varphi$ and $\varphi^2$ are not allowed to be generated. 

Therefore, in the perturbative argument, we can conclude that any relevant or marginal operators 
except nondynamical operators do not appear radiatively, 
meaning that no fine tuning is required 
to take the continuum limit.

\subsection{Matrix String theory} 
\label{sec:MST}
The mass-deformed $\cN=(8,8)$ SYM in two dimensions can be obtained 
from the constructed lattice theory around the trivial minimum $C=\phi_\pm=0$ 
as seen in the previous section. 
Since $M$ is a soft mass breaking 16 SUSYs to $Q_\pm$, undeformed theory, which is 
nothing but the IIA matrix string theory \cite{MatrixString}, can be defined by 
turning off $M$ after the continuum limit.

\setcounter{equation}{0}
\section{4d $\cN=4$ SYM} 
\label{sec:4dN=4}
In this section, we discuss a scenario to obtain four-dimensional $\cN=4$ SYM from the lattice formulation 
given in the previous section. 

Let us consider the lattice theory expanded around the minimum of $k$-coincident fuzzy $S^2$ given 
by (\ref{fuzzy sphere solution}) 
with 
\be
L_a=L_a^{(n)}\otimes \id_k \quad (a=1,2,3) \quad \mbox{and} \quad N=nk.
\label{k_FS2}
\ee 
$L_a^{(n)}$ are $SU(2)$-generators of an $n(=2j+1)$-dimensional irreducible representation 
of spin $j$. 

First, we take the continuum limit of the two-dimensional lattice theory. 
Then, we obtain four-dimensional 
$\cN=4$ $U(k)$ SYM on ${\bf R}^2 \times (\mbox{Fuzzy }S^2)$ 
with 16 SUSYs broken to $Q_\pm$ by $M$~\footnote{Due to the infinite volume of ${\bf R}^2$, 
tunnelling among discrete minima of various fuzzy sphere solutions is suppressed 
to stabilize each fuzzy sphere background.  
}.      
The fuzzy $S^2$ has the radius $R=3/M$, and its noncommutativity (fuzziness) is characterized by the parameter 
$\Theta= \frac{18}{M^2n}$. UV cutoff in the $S^2$ directions is set 
at $\Lambda= \frac{M}{3}\cdot 2j$. 
These properties of the fuzzy $S^2$ are seen by doing a similar calculation as presented in 
Refs.~\citen{Iso:2001mg,Maldacena:2002rb,Ishii:2008ib}.  
In particular, momentum modes of a field, say $B_{\bA}$, on two dimensions are expanded further by 
fuzzy spherical harmonics: 
\be
\tilde{B}_{\bA}(q) = \sum_{J=0}^{2j}\sum_{m=-J}^J \hat{Y}^{(jj)}_{J\,m}\otimes b_{\bA, \,J\,m},
\label{expand_B}
\ee  
corresponding to the expression (\ref{k_FS2}). 
The fuzzy spherical harmonic $\hat{Y}^{(jj)}_{J\,m}$ is an $n\times n$ matrix whose elements are given by  
Clebsch-Gordon (C-G) coefficients as~\cite{Ishii:2008ib}
\be
\hat{Y}^{(jj)}_{J\,m} = \sqrt{n}\sum_{r,r'=-j}^j(-1)^{-j+r'}C^{J\,m}_{j\,r\,j\,-r'}\,|j\,r\ket\bra j\,r'|
\ee
with an orthonormal basis $|j\,r\ket$ representing $L^{(n)}_a$ in the standard way: 
\bea
\left(L_1^{(n)}\pm iL_2^{(n)}\right)\,|j\,r\ket & = & \sqrt{(j\mp r)(j\pm r+1)}\,|j\,r\pm 1\ket, \nn \\
L_3^{(n)}\,|j\,r\ket & = & r\,|j\,r\ket, 
\eea
and the modes $b_{\bA,\,J\,m}$ are $k\times k$ matrices. It is seen that the fuzzy spherical harmonics are 
eigen-modes of the Laplacian on the fuzzy $S^2$: 
\be
\sum_{a=1}^3\left(\frac{M}{3}\right)^2[L^{(n)}_a, [L^{(n)}_a, \hat{Y}^{(jj)}_{J\,m}]] 
= \left(\frac{M}{3}\right)^2J(J+1)\hat{Y}^{(jj)}_{J\,m}, 
\ee
giving the rotational energy with the angular momentum $J$ on the sphere of the radius $R=3/M$.   
The UV cutoff $\Lambda= \frac{M}{3}\cdot 2j$ can be read off from the upper limit of the sum of $J$ 
in the expansion (\ref{expand_B}). 
The fuzzy $S^2$ is a two-dimensional noncommutative (NC) space, which is analogous to the phase space of 
some one-dimensional quantum system, 
and the noncommutativity $\Theta$ to the Planck constant $\hbar$. 
The quantum phase space is divided into small cells of the size $2\pi \hbar$, whose number is equal to 
the dimension of the Hilbert space. 
Correspondingly, the area of the $S^2$ is divided into $n$ cells of the size $2\pi\Theta$: 
\be
4\pi R^2 = n\cdot 2\pi \Theta,  
\ee
leading to the value $\Theta=\frac{18}{M^2n}$.    

Notice, differently from the two-dimensional case, 
it is not clear whether the SUSY breaking by $M$ is soft, because $M$ appears not only 
in mass terms in the action  
but also in the geometry of the fuzzy $S^2$. 
Let us proceed assuming that the breaking is soft~\footnote{
The assumption is plausible from the viewpoint of 
the mapping rule between matrix model and Yang-Mills theory 
on noncommutative space~\cite{Aoki:1999vr}.}.  
We will give some argument later for 
the validity of the assumption.  

Next, we take successive limits by following the two steps: 
\begin{itemize}
\item {\bf Step 1}: 
Take large $n$ limit with $\Theta$ and $k$ fixed. Namely, $M\propto n^{-1/2}\to 0$ 
and $\Lambda\propto n^{1/2} \to \infty$. 
\item {\bf Step 2}: 
Send $\Theta$ to zero. 
\end{itemize}

\subsection{Step 1} 
At the step 1, the fuzzy $S^2$ is decompactified to the NC Moyal plane ${\bf R}^2_\Theta$. 
{}From the assumption, the theory becomes $\cN=4$ $U(k)$ SYM on ${\bf R}^2\times {\bf R}^2_\Theta$ 
with {\em 16 SUSYs restored}. The gauge coupling constant of the four-dimensional theory is given in the form 
\be
g_{4d}^2 =2\pi \Theta g_{2d}^2. 
\label{4d_2d_coupling}
\ee 
In the limit, the expansion (\ref{expand_B}) by the fuzzy spherical harmonics can be 
essentially transcribed to  
the one by plane waves on ${\bf R}^2_{\Theta}$:  
\be
\tilde{B}_{\bA}(q) = \int\frac{d^2\tilde{q}}{(2\pi)^2}\,e^{i\tilde{q}\cdot\hat{x}}\otimes \tilde{b}_{\bA}(\bq), 
\ee
where $q$ and $\tilde{q}$ are two-momenta on ${\bf R}^2$ and ${\bf R}^2_{\Theta}$ respectively,  
the position operator $\hat{x}=(\hat{x}_1, \hat{x}_2)$ on ${\bf R}^2_{\Theta}$ satisfies 
$[\hat{x}_1, \hat{x}_2]=i\Theta$, and $\bq\equiv (q,\tilde{q})$ represents a four-momentum. 
The modes $\tilde{b}_{\bA}(\bq)$ in the four-dimensional space are $k\times k$ matrices.  
It is easy to calculate the inner product between plane waves on ${\bf R}^2_{\Theta}$:   
\be
\Tr\left(e^{i\tilde{p}\cdot\hat{x}}e^{i\tilde{q}\cdot\hat{x}}\right) 
= \frac{2\pi}{\Theta}\,\delta^2(\tilde{p}+\tilde{q}),  
\ee
which leads to the $\Theta$-dependence of the relation (\ref{4d_2d_coupling}). 

Let us discuss radiative corrections in four-dimensional SYM on 
${\bf R}^2 \times (\mbox{Fuzzy }S^2)$. 
We give an argument below that there is no radiative correction which prevents from the 
full 16 SUSYs being restored after the step 1. 
  
In quantum field theory defined on NC space with a constant noncommutativity, there are 
two kinds of Feynman diagrams. 
One is planar diagrams. They have no NC phase factors depending on loop momenta, 
and their behavior is 
the same as that in the corresponding theory on the ordinary space~\cite{Eguchi:1982ta}. 
The other is nonplanar diagrams. They have NC phase factors, which improve the UV behavior of the diagrams. 
But, when some of the NC phases vanish in the infra-red (IR) region of external or loop momenta, 
singularities may arise, whose origin is 
the UV singularities in the corresponding theory on the ordinary space  
(UV/IR mixing)~\cite{Minwalla:1999px}. 
Therefore, we can say that 
UV behavior of planar and nonplanar diagrams in the theory on NC space is not worse 
than that in the corresponding theory on 
the ordinary space. 
Let us consider the superficial degree of UV divergences of Feynman diagrams in ordinary 
four-dimensional gauge theory: 
\be
D=4-E_B-\frac32 E_F, 
\ee
where $E_B$ ($E_F$) is the number of the external lines of bosons (fermions). 
In our case, the divergence of $D=3$, that is from $E_B=1$, is absent 
since the operator $\varphi$ is forbidden 
by $Q_\pm$ SUSYs as in the two-dimensional case. Thus, the possible most severe divergences are of the 
degree $D=2$. 
The leading $\Lambda^2$ terms are expected to cancel each other by 16 SUSYs 
under the assumption that $M$ is soft.  
For radiative corrections to gauge invariant observables, 
divergences possibly originate from the mass deformation, whose behavior is expected as~\footnote{
We should note that the behavior (\ref{Mpq}) is not valid for gauge-dependent divergences 
which can be absorbed into wave function renormalization. 
(\ref{Mpq}) is derived based on UV finiteness of the undeformed four-dimensional $\cN=4$ 
SYM~\cite{Mandelstam:1982cb,Erickson:2000af}, 
where the finiteness holds except such divergences. 
}   
\be
M^p\left(\ln\frac{\Lambda}{M}\right)^q=\cO\left(M^p(\ln n)^q\right) 
\qquad 
 (p=1,2, \, q=1,2,\cdots). 
 \label{Mpq}
\ee
However, such terms disappear in the limit of the step 1. 

Hence, there appears no radiative correction preventing restoration of the full SUSYs 
after the step 1, leading to $\cN=4$ $U(k)$ SYM on ${\bf R}^2\times {\bf R}^2_\Theta$ 
with 16 supercharges.   

\subsection{Step 2} 
In four-dimensional $\cN=4$ SYM on NC space, the commutative limit ($\Theta\to 0$) is believed to be 
smooth~\cite{Hashimoto:1999ut,Matusis:2000jf},  
that is, desired $\cN=4$ $U(k)$ SYM on usual flat ${\bf R}^4$ should be obtained with no fine tuning 
after the step 2. 

\subsection{Check of the scenario} 
As a check of the scenario presented in the above, we computed 1-loop radiative corrections to 
scalar kinetic terms of $B_{{\bA}=1,2}$. Although details are presented in a separate publication~\cite{HMSS}, 
contribution from planar diagrams to the kinetic terms in the 1-loop effective action finally becomes    
\be
\frac{1}{g_{4d}^2}\int\frac{d^4\bq}{(2\pi)^4}\,\sum_{{\bA}=1,2}\tr_k
\left[\bq^2\,\tilde{b}^{(R)}_{\bA}(-\bq)\tilde{b}^{(R)}_{\bA}(\bq)\right]
\left\{1+\frac{g_{4d}^2\,k}{4\pi^2}\left(-\frac12\ln\frac{\bq^2}{\mu_R^2} +1\right) +\cO(g_{4d}^4)\right\} 
\label{Seff_1-loop}
\ee
after the limit of the step 1. 
$\tilde{b}^{(R)}_{\bA}(\bq)$ are renormalized 
momentum modes which are related to 
the modes 
$\tilde{b}_{\bA}(\bq)$ 
of the bare fields $B_{\bA}$ 
as  
\be
\tilde{b}^{(R)}_{\bA}(\bq)= \left(1+\frac{g_{4d}^2\,k}{4\pi^2}\ln\frac{\Lambda}{\mu_R}\right)^{1/2} \,\tilde{b}_{\bA}(\bq),
\label{wf_ren}
\ee
where $\mu_R$ is the renormalization point~\footnote{
Although four-dimensional $\cN=4$ SYM is said to be 
UV finite, divergence of the wave function renormalization can appear as 
a gauge artifact~\cite{Mandelstam:1982cb}. In fact, a modified light-cone gauge fixing in 
Refs.~\citen{Mandelstam:1982cb} leads to 
no UV divergence even in the part concerning the wave function renormalization, 
differently from the 1-loop computation in Feynman gauge fixing~\cite{Erickson:2000af}. 
The point is that even if UV divergences appear in radiative corrections, 
all of them can be absorbed by rescaling the fields~\cite{Erickson:2000af}. 
We adopted a Feynman-like gauge fixing in the calculation. 
}.   
The logarithmic nonlocal term 
in (\ref{Seff_1-loop}) has a definite physical meaning contributing 
to anomalous scaling dimension of the operator. 
Since the result does not depend on $\Theta$, the limit of the step 2 is trivial. 

Note that four-dimensional rotational symmetry is restored in (\ref{Seff_1-loop}), 
which can be regarded as an evidence 
of the restoration of the full 16 SUSYs and of the softness of $M$. 
Furthermore, we found that nonplanar contribution is essentially the same as the planar contribution, 
supporting the smoothness of the commutative limit $\Theta\to 0$. 

\setcounter{equation}{0}
\section{Discussions} 
\label{sec:conclusion}
We constructed a lattice formulation of two-dimensional $\cN=(8,8)$ SYM with a mass deformation,  
which preserves two supercharges. It serves a basis of 
nonperturbative investigation of the IIA matrix string theory.  
Also, it gives an intriguing scenario to obtain four-dimensional $\cN=4$ $U(k)$ SYM with arbitrary $k$, 
requiring no fine tuning.   

It is interesting to extend such construction to theories coupled to fundamental matters.   
Although it is difficult to introduce fundamental fields directly, 
bi-fundamental fields can easily be incorporated. 
For instance, let us start with a two-dimensional system with SUSYs, which 
is obtained by dimensional reduction from the corresponding theory in four dimensions. 
The action $S_0=S_{0, g}+S_{0,g'}+S_m$ is   
\begin{itemize}
\item
$S_{0,g}$ is the action of $U(N)$ SYM with gauge field $A_\mu$ and adjoint matters $X_I$ 
($I=1, \cdots, \ell_a$): 
\be
S_{0, g}=\frac{1}{g_{2d}^2}\int d^2x\,\tr_N\left[F_{12}^2+(D_\mu X_I)^2 -\frac12[X_I,X_J]^2 
+ \mbox{(fermions)}\right] 
\label{action_0g}
\ee
with $D_\mu X_I =\partial_\mu X_I+i[A_\mu, X_I]$.  
\item
$S_{0,g'}$ is the action of $U(N')$ SYM with gauge field $A'_\mu$ and adjoint matters $X'_I$ 
($I=1, \cdots, \ell_a$):
\be
S_{0, g'}=\frac{1}{(g'_{2d})^2}\int d^2x\,\tr_{N'}\left[(F'_{12})^2+(D_\mu X'_I)^2 -\frac12[X'_I,X'_J]^2 
+ \mbox{(fermions)}\right] 
\ee
with $D_\mu X'_I =\partial_\mu X'_I+i[A'_\mu, X'_I]$. 
It is essentially the same as (\ref{action_0g}) except the change $g_{2d}\to g'_{2d}$, $N \to N'$.   
\item
$S_m$ is the action of $U(N)\times U(N')$ bi-fundamental matters $\Phi_i$ ($i=1, \cdots, \ell_f$) 
coupled to the above two systems: 
\be
S_m = \int d^2x\,\tr_{N'}\left[(D_\mu \Phi_i)^\dagger D_\mu \Phi_i 
+\left(X_I\Phi_i-\Phi_iX'_I\right)^\dagger \left(X_I\Phi_i-\Phi_iX'_I\right) + \mbox{(fermions)}\right]
\ee
with $D_\mu \Phi_i = \partial_\mu \Phi_i +iA_\mu\Phi_i-i\Phi_iA'_\mu$. 
\end{itemize}
We consider the situation that both of $S_{0, g}$ and $S_{0,g'}$ allow a deformation by mass $M$ preserving 
some SUSYs as discussed in section~\ref{sec:continuum}, 
and that deformed actions $S_g$ and $S_{g'}$ have classical 
solutions of $k$- and $k'$-coincident fuzzy $S^2$: 
\bea
& & X_a=\frac{M}{3}\,L_a^{(n)}\otimes \id_k \qquad  (N=nk), \nn \\ 
& & X'_a=\frac{M}{3}\,L_a^{(n)}\otimes \id_{k'} \qquad (N'=nk'),
\label{FS2_kk'}
\eea  
with all the other fields nil, respectively. 
(We labelled the index $I$ so that scalars with $I=1,2,3$ satisfy the fuzzy $S^2$ configurations.)   
Then, for $k$ and $k'$ general, vanishing the bi-fundamental fields gives the minima of the zero total 
action $S\equiv S_g +S_{g'}+S_m=0$.  
Expanding $S$ around the background (\ref{FS2_kk'}) 
leads to two systems of gauge and adjoint fields with gauge groups $U(k)$ and $U(k')$,  
which are coupled by $U(k)\times U(k')$ bi-fundamental matters. 
They are defined on ${\bf R}^2\times$ (Fuzzy $S^2$), analogous to the situation seen in 
section~\ref{sec:4dN=4}. 
Finally, after turning off the coupling $g'_{2d}$, 
we obtain the system of $U(k)$ gauge and adjoint fields 
coupled to $k'\ell_f$ fundamental matters 
(with $U(k')$ gauge and adjoint fields becoming free and decoupled) on 
${\bf R}^2\times$ (Fuzzy $S^2$). 
Therefore, if the system of the action $S$ can be realized on lattice, 
and if successive limits analogous to those discussed in section~\ref{sec:4dN=4} can be taken safely, 
it is expected to obtain the desirable quantum system on ${\bf R}^4$.  
(Similar construction using the Eguchi-Kawai 
equivalence can be found in Ref.~\citen{Hanada:2009hd}.)  

Using our formalism, many interesting theories will be realized on computer.  
We expect new insights into nonperturbative dynamics of supersymmetric theories 
will be obtained in near future.

\section*{Acknowledgements}
The authors are deeply grateful to Hiroshi~Suzuki for his joining to their loop calculation. 
They would like to thank Ofer~Aharony, Adi~Armoni, Masafumi~Fukuma, Hikaru~Kawai, Jun~Nishimura, 
Adam~Schwimmer, Hidehiko~Shimada, Asato~Tsuchiya, 
Mithat~\"{U}nsal and Kentaroh~Yoshida for stimulating 
discussions and comments.     
F.~S. would like to thank Weizmann Institute of Science 
for hospitality during his stay.  
S.~M. would like to thank Jagiellonian University 
for hospitality during his stay. 
The work of S.~M. is supported in part by Keio Gijuku Academic Development Funds, and   
the work of F.~S. is supported in part by Grant-in-Aid
for Scientific Research (C), 21540290.

\appendix
\section{Notations} 
\label{app:notation}
We give the relation between field variables in (\ref{S0}) and those in (\ref{Q^{(0)}-closed form}), 
(\ref{4d N=4 continuum action}). 

For scalars, 
\be
B_1 = -X_5, \quad B_2=X_6, \quad B_3 = X_7, \quad C = 2X_8, \quad \phi_\pm = X_9 \pm iX_{10}. 
\ee
For fermionic variables, 
\bea
\Psi & = & U_{16}\Psi^{(0)}, \nn \\
(\Psi^{(0)})^T & \equiv &  
\left(\rho_{+3},\rho_{+4}, \psi_{+2}, \psi_{+1}, -\chi_{+1}, \chi_{+2}, \chi_{+3}, \frac12\eta_+, \right. 
\nn \\
& & \hspace{3mm} \left.   
\rho_{-3},\rho_{-4}, \psi_{-2}, \psi_{-1}, -\chi_{-1}, \chi_{-2}, \chi_{-3}, \frac12\eta_-\right),
\eea
where $U_{16}$ is a $16\times 16$ unitary matrix of the form 
\be
U_{16} = \frac12 \begin{pmatrix}  & \cA_{12} & \cA_{13} &   \\ \cA_{21} &  &  & \cA_{24} \end{pmatrix}
\ee 
with 
\bea
& & \cA_{12} \equiv \begin{pmatrix} 0 & 0 & -i & 1 \\ i & 1 & 0 & 0 \\ 0 & 0 & -1 & -i \\ 1 & -i & 0 & 0 \\
i & -1 & 0 & 0 \\ 0 & 0 & i & 1 \\ 1 & i & 0 & 0 \\ 0 & 0 & 1 & -i \end{pmatrix}, \qquad 
\cA_{13} \equiv \begin{pmatrix} 0 & 0 & -1 & i \\ -1 & i & 0 & 0 \\ 0 & 0 & -i & -1 \\ -i & -1 & 0 & 0 \\
-1 & -i & 0 & 0 \\ 0 & 0 & 1 & i \\ -i & 1 & 0 & 0 \\ 0 & 0 & i & -1 \end{pmatrix}, \nn \\
& & \cA_{21} \equiv \begin{pmatrix} 0 & 0 & i & 1 \\ i & 1 & 0 & 0 \\ 0 & 0 & -1 & i \\ -1 & i & 0 & 0 \\
i & -1 & 0 & 0 \\ 0 & 0 & -i & 1 \\ -1 & -i & 0 & 0 \\ 0 & 0 & 1 & i \end{pmatrix}, \qquad 
\cA_{24} \equiv \begin{pmatrix} 0 & 0 & -1 & i \\ 1 & -i & 0 & 0 \\ 0 & 0 & i & 1 \\ -i & -1 & 0 & 0 \\
1 & i & 0 & 0 \\ 0 & 0 & 1 & i \\ -i & 1 & 0 & 0 \\ 0 & 0 & -i & 1 \end{pmatrix}. 
\eea

Also, the explicit form of the gamma matrices we used is 
\bea
\gamma_2 & = & -i \id_2 \otimes \id_2 \otimes \id_2 \otimes \sigma_3, \nn \\
\gamma_3 & = & -i \id_2 \otimes \id_2 \otimes \id_2 \otimes \sigma_1, \nn \\
\gamma_4 & = & +i \id_2 \otimes \id_2 \otimes \sigma_2 \otimes \sigma_2, \nn \\
\gamma_5 & = & -i \sigma_2 \otimes \sigma_3 \otimes \sigma_3 \otimes \sigma_2, \nn \\
\gamma_6 & = & +i \sigma_2 \otimes \id_2 \otimes \sigma_1 \otimes \sigma_2, \nn \\
\gamma_7 & = & -i \sigma_2 \otimes \sigma_1 \otimes \sigma_3 \otimes \sigma_2, \nn \\
\gamma_8 & = & -i \sigma_1 \otimes \sigma_2 \otimes \sigma_1 \otimes \sigma_2, \nn \\
\gamma_9 & = & +i \id_2 \otimes \sigma_2 \otimes \sigma_3 \otimes \sigma_2, \nn \\
\gamma_{10} & = & +i \sigma_3 \otimes \sigma_2 \otimes \sigma_1 \otimes \sigma_2
\eea
with $\sigma_a$ ($a=1,2,3$) being the Pauli matrices. 
 

\end{document}